# Deep Multi-Scale Representation Learning with Attention for Automatic Modulation Classification

1st Xiaowei Wu
*School of Information and Communication*
*National University of Defense Technology*
Wuhan, China
wxwgo@qq.com

2nd Shengyun Wei*
*School of Information and Communication*
*National University of Defense Technology*
Wuhan, China
junyun1002@126.com

3rd Yan Zhou
*Schoole of Information and Communication*
*National University of Defense Technology*
Wuhan, China
2329451129@qq.com

*Abstract*—Currently, deep learning methods with stacking small size convolutional filters are widely used for automatic modulation classification (AMC). In this report, we find some experienced improvements by using large kernel size for convolutional deep convolution neural network based AMC, which is more efficient in extracting multi-scale features of the raw signal I/Q sequence data. Also, Squeeze-and-Excitation (SE) mechanisms can significantly help AMC networks to focus on the more important features of the signal. As a result, we propose a multi-scale feature network with large kernel size and SE mechanism (SE-MSFN) in this paper. SE-MSFN achieves state-of-the-art classification performance on the public well-known RADIOML 2018.01A dataset, with average classification accuracy of 64.50%, surpassing CLDNN by 1.42%, maximum classification accuracy of 98.5%, and an average classification accuracy of 85.53% in the lower SNR range 0dB to 10dB, surpassing CLDNN by 2.85%. In addition, we also verified that ensemble learning can help further improve classification performance. We hope this report can provide some references for developers and researchers in practical scenes.

*Keywords*—AMC, Large kernel size, SE-MSFN, Ensemble learning

## I. Introduction

In wireless communication systems, AMC is a necessary step between signal detection and demodulation, and its main task is to identify the modulation type of the received signal by effective means, so as to lay the foundation for subsequent signal demodulation and information acquisition. When implementing electronic countermeasures, recognizing the modulation type of the enemy signal is a prerequisite for effective radio jamming and spoofing [1]. In the field of intelligence reconnaissance, the modulation type of a captured signal is first recognized in order to quickly obtain the information it conveys. In cognitive radio, modulation type recognition is the cornerstone of cognitive radio's adaptive modulation and demodulation capabilities, which can sense and learn from the environment and adapt accordingly [2]. In the process of spectrum monitoring, modulation information is an important factor in deciding whether to follow radio rules [3]. With the development of intelligent communication systems, AMC will also play an even more important role.

Traditional AMC methods are broadly classified into two categories, likelihood-based (LB) methods [4], [5] and signal feature-based (FB) methods [6], [7]. LB methods are statistical-based methods that require the calculation of the maximum likelihood function for each modulation modes. The FB method first extracts the expert features of the signal and then feeds them into traditional machine learning classifiers (e.g. SVM, DTree, NaiveByaes, and KNN) for classification, but traditional machine learning algorithms do not have strong generalization capabilities and cannot face complex modulation classification tasks. In recent years, deep learning (DL) has shown excellent performance in fields such as CV (Computer Vision), speech recognition and NLP (Natural Language Processing) [8]-[10], as a result, more and more researchers are applying this approach to AMC as well. A simple CNN was first applied to solve the AMC problem [11], which takes the in-phase and quadrature (I/Q) sequence data of the raw signal as input directly, uses the CNN to extract and learn the signal features autonomously, and then recognizes the modulation type by a classifier, and the classification performance outperforms traditional machine learning methods. Then, the deep residual learning [12] was proposed to solve the degradation problem of deep neural networks (DNNs), for which researchers proposed two modulation classification models based on ResNet and VGG [13], which achieved better classification performance than CNN on the RADIOML 2018.01A dataset. Meanwhile, considering the timing characteristics of the raw signal I/Q sequence data, the recurrent neural network (RNN), which is good at processing sequence data, was also applied to solve the AMC problem [14]. Combined CNN and LSTM [15] to get the Inception-LSTM and Res-LSTM [16], in which the CNN module of the model can extract the high-dimensional complex signal data into low-dimensional feature vectors to eliminate redundancy, and the LSTM module can make the network remember the low-dimensional data that require long time to remember and forget the unimportant information through the dependency relationship between adjacent sequence data. Simulation results show that the classification accuracy was improved but the prediction time was longer compared to the simple CNN. Moreover, researchers also combined ResNet with GRU [17] to get the GrrNet [18], which has a similar structure to the LSTM, but with fewer parameters and easier convergence than the LSTM. The classification mechanism of GrrNet is similar to the Inception-LSTM and Res-LSTM [16], except that the input to GrrNet is the signal amplitude and phase (A/P) feature data, and achieves better classification performance at high SNR. Since the RNN-based modulation classification model only considers the temporal features of the received signal sequence data (I/Q and A/P), so the model is less robust against noise. To solve this problem, an MSN modulation classification model based on signal multi-scale feature fusion was proposed [19], which uses multiple multi-scale fusions to fuse the multi-scale feature of the raw signal I/Q sequence data together, which can better learn the local differences and timing characteristics of different modulated signals, and the classification accuracy is greatly improved and reflects better robustness to varying SNR environments. However, the classification performance of MSN at low SNR is not good enough. In conclusion, the existing DL-based AMC methods have a lot of room for improvement in classification performance at lower SNR. There are two reasons for this: (1) The backward and forward correlations and dependencies of the raw signal I/Q sequence data are ignored, which is extremely important for classification between similar modulation modes. (2) The dependence of the network on different convolutional feature channels is ignored, and the different convolutional feature channels are not enhanced or suppressed according to their importance.

Therefore, a multi-scale convolutional neural network AMC algorithm based on large kernel size and channel self-attentiveness mechanism is proposed in this paper. i.e. SE-MSFN, which can effectively focus on the backward and forward correlations and dependencies of sequence data when extracting multi-scale feature of the raw signal I/Q sequence data, and the SE mechanism can enhance the important convolutional feature channels and suppress the unimportant convolutional feature channels during the training process, so as to obtain the multi-scale features of the signal more efficiently. The public challenging RADIOML 2018.01A dataset was used to test the performance of SE-MSFN. The simulation results show that the classification performance of SE-MSFN is outperforming other methods. In addition, we verified that the ensemble learning approach can further improve the classification accuracy.

The remaining sections are organized as follows. Section II describes the related work on AMC. Section III presents AMC system model. Section IV describes the SE-MSFN algorithm. Section V evaluates the performance of SE-MSFN. Finally, Section VI draws the conclusion.

## II. RELATED WORK

Deep convolutional neural networks have powerful learning and generalization capabilities, not only to automatically extract features of modulated signals, but also to effectively deal with the challenges of unknown environments [20]-[23]. DL-based AMC methods are usually divided into two stages, with the first stage pre-processing the received signal and converting it into different representational forms before feeding it into the network for classification. In the signal pre-processing stage, signal representation can be broadly classified into two categories. The first category is expert feature-based signal representation [24]-[27], which first extracts the high-order accumulation, spectra, cyclic characteristics or other statistical features of the received signal, to convert the received signal into features in the time or frequency domains to obtain a signal representation of one or more fused features, followed by classification of the modulation modes. This method requires the manual selection of different expert features for different modulation modes, with a high operational complexity and poor generalization capabilities, because the different expert feature are only valid for classification of several modulation modes. The second category is image-based signal representation [28]-[31], which first converts received signals into feature forms such as constellation diagrams, eye diagrams and feature point diagrams, and then inputs them into the classifier. Although the method takes advantage of the powerful capabilities of DNNs in the field of CV, the classification process is complex and less robust to noise. Particularly in the case of high noise interference, the image features of different modulation modes are extremely similar, making it difficult for the network to achieve better classification performance. In contrast, the direct use of the raw signal I/Q sequence data as input eliminates the need for data pre-processing and greatly reduces the computational complexity. Additionally, this approach allows the model to deal with different AMC tasks in a variety of unknown environments more efficiently.

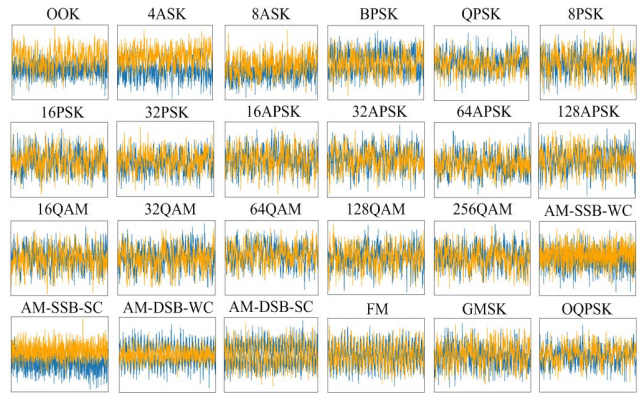

Fig. 1. I/Q time domain examples of 24-modulation modes at 0dB SNR.

Fig. 1 shows the I/Q time domain view of the 24 modulated signals of public well-known RADIOML 2018.01A dataset at 0dB. As we can see from the Fig. 1, it is very difficult to distinguish them under strong interference of noise if the AMC network only extracts features at a single scale of the signal. Only if the AMC network can extract multi-scale feature maps of the signal and fuse them at multiple scales can learn the typical features of different modulation modes efficiently, which can improve the feature extraction of the similar modulated signals at low SNR and the classification performance will be greatly improved. This method is inspired by the High Resolution Network (HRNet) [32], [33], which was first proposed for human pose estimation and achieves excellent estimation performance by concatenating high-resolution to low-resolution convolutions in parallel to obtain a rich high-resolution representation and gradually adding the low-resolution feature maps to the high-resolution ones. Moreover, SE attention mechanism [34] is employed to enhance and suppress features based on their interdependence between channels. By fusing SE block into MSFN, the attention mechanism can help the network learn the more important information in multi-scale feature maps of the signals.

## III. SYSTEM MODEL

In wireless communication systems, AMC is a multi-classification problem for the modulation modes of the received signal, which determines the type of modulation based on the characteristics of the received signal sequence. In general, the original received signal can be expressed as:

$$s(k) = Ae^{j(2\pi f_0 kT + \theta_k)} \cdot \sum_{-\infty}^{\infty} x(n)h(kT - nT + \varepsilon_T T) + n(k) \quad (1)$$

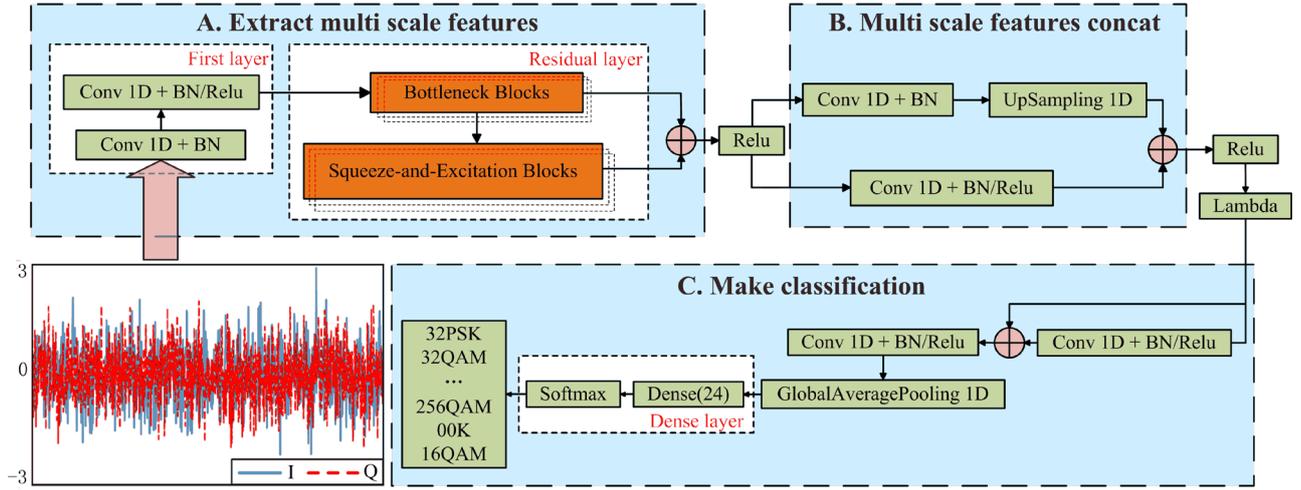

Fig. 2. The architecture of SE-MSFN.

Where $n(k)$ denotes additive white Gaussian noise (AWGN), $x(n)$ denotes signal symbol sequence, $A$ denotes amplitude fading factor, $f_0$ denotes frequency carrier frequency offset, $\theta_k$ denotes phase jitter, $T$ denotes symbol spacing, $h(\cdot)$ denotes residual channel effects, and $\varepsilon_T$ denotes timing error. Then the in-phase and quadrature components of $s(k)$ can be expressed respectively as:

$$\begin{cases} I_k = real(s(k)) \\ Q_k = imag(s(k)) \end{cases} \quad (2)$$

Further, assume that a receive vector consisting of N sample points of the receive signal I/Q sequence $\vec{S} = [I_0, I_1, ..., I_{N-1}, Q_0, Q_1, ..., Q_{N-1}]$, the candidate set of modulation modes is $\{M_0, M_1, ..., M_{L-1}\}$, the final modulation type classification task can then be expressed as:

$$M_i = f(\vec{S}), i \in [0, L-1] \quad (3)$$

Also, to represent the effect of noise on the signal, the SNR can be defined as:

$$SNR = 10 \log_{10}(\frac{p_s}{p_n}) \quad (4)$$

Where $p_s$ and $p_n$ denote the effective power of the signal and noise, respectively. This is one of the important factors affecting the performance of the AMC.

IV. METHODOLOGY

Fig. 2 shows the general architecture of the SE-MSFN model, which takes the raw signal I/Q sequence data directly as input. Firstly, a large size convolutional filter is adopted, which can better focus on the backward and forward correlation and dependency of the data in the process of extracting features. In order to extract the multi-scale feature of the signal I/Q sequence data efficiently, the network joined with the SE module, which can help adaptively adjust the learning weights of different convolutional feature channels according to the loss during the training process and the more important features of the I/Q sequence data will be extracted. And then, the fusing multi-scale features of the I/Q sequence data are fed into the classifier for classification. In addition, we have integrated several weak models using an ensemble learning approach to further improve the average classification accuracy. In the following we will explain the details of the architecture of SE-MSFN and theory of the proposed method.

*A. Large kernel Size*

The raw signal I/Q sequence data is a kind of time sequence data, which is different from image data in CV, and is more similar to text data in NLP. For classification tasks, the backward and forward correlation and long-term dependency features of sequence data are more important than edge features and local features. Convolutional filters with large kernel size are able to focus better on the front-to-back correlation and global features of the I/Q sequence data for its larger perceptual field, more in-depth multi-scale features of the I/Q sequence data can be extracted effectively. However, it has long been considered that convolutional filters with large kernel size are inefficient and ineffective for feature extraction, and small size convolutional filters are more popular with researchers.

*B. Squeeze and Excitation Blocks*

SE blocks aims to explicitly model the interdependence relationship between feature channels, where importance of different channels are learnt automatically for feature enhancement and suppression [34]. Fig. 3 shows the structure of the SE Module, mainly consisting of a layer of one-dimensional global average pooling and two layers of fully connected layer functions, with Relu and Sigmoid as activation functions for the fully connected layers, respectively.

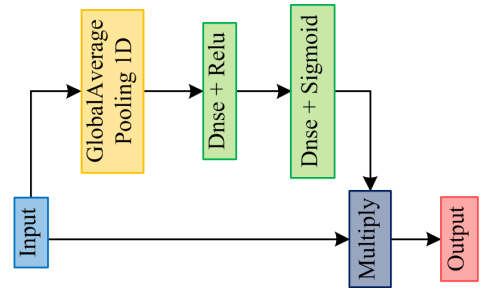

Fig. 3. Structure of SE Module

As shown in Fig. 4. The SE Module has two important operations: squeeze and excitation. Taking the input one-dimensional I/Q data as an example, the squeeze operation compresses the extracted global feature map information of the I/Q data into a channel descriptor by one-dimensional global averaging pooling, which can be represented as:

$$z_C = F_{sq}(u_c) = \frac{1}{1 \times W} \sum_{j=1}^{W} u_C(1,j) \quad (5)$$

Where $C$ denotes the number of channels, $z_C$ denotes the $C$th channel description element of the output of the squeeze operation and the dimension is $1 \times C$, and $u_C$ denotes the $C$th feature map of the feature and the dimension is $1 \times W$.

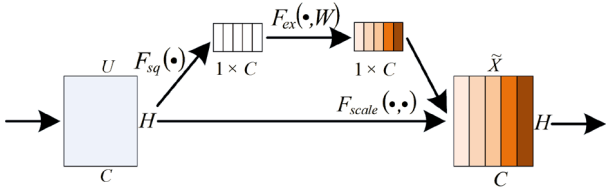

Fig. 4. The details of the one-dimensional SE mechanism

After obtaining the global feature description, the excitation operation first feeds it into the two fully connected layers, then the result is restricted to the range of 0 to 1 by the sigmoid function, and the dependence of the network on each convolutional channel is then obtained, a process that can be represented as:

$$s = F_{ex}(z, W) = \sigma(g(z, W)) = \sigma(W_2 \delta(W_1 z)) \quad (6)$$

Where $W_1$, $W_2 \in \mathbb{R}^{C/r \times C}$ represents a fully connected layer for raising or lowering dimensionality, $r$ denotes compression ratio, $\delta$ denotes ReLU activation function, $\sigma$ denotes sigmoid function. Finally, the resulting weights for each channel are multiplied by the Reshape and Multiply functions corresponding to the original feature maps. This process can be represented as:

$$\widetilde{x}_C = F_{scale}(u_C, s_C) = s_C \cdot u_C \quad (7)$$

Where $\widetilde{x}_C$ denotes the feature maps after loading the weights, and the whole SE operation is completed, allowing the network to strengthen the important feature channels and suppress the less important ones.

### C. Model structure

Fig. 2 shows the architecture of SE-MSFN, consisting of Part A, Part B, and Part C. Considering that the I/Q sequence data are one-dimensional time sequence data, one-dimensional convolution filters (Conv1D) are used for all convolution layers in order to reduce the computational complexity.

Part A consists of a First layer and a Residual layer, using the raw signal I/Q sequence data as direct input. The First layer consists of two layers of Conv1D, both with convolutional filters of 32, kernel size of 9×1 and stride of 1, allowing a wide range of low-order features to be extracted. Residual layer consisting of Bottleneck Blocks and SE Module, the structure of Bottleneck Blocks is shown in Fig. 5, which consists of three layers of Conv1D with convolutional filters of 32, the first and third layer with kernel size of 1×1 and stride of 1, which are used to perform up-dimensioning or down-dimensioning of the feature maps, and the second layer with kernel size of 9×1 and strides of 2. The SE-MSFN model has four layers of stacked Bottleneck Blocks, and each Bottleneck Block is connected to the SE Module as a residual to obtain multi-scale sequence feature maps of the raw signal I/Q sequence data.

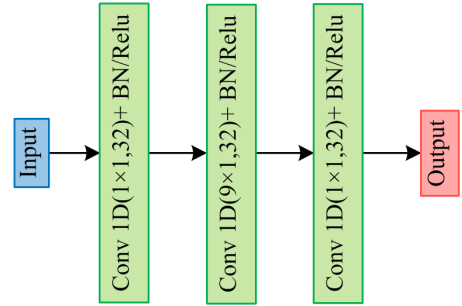

Fig. 5. Structure of bottleneck blocks, the text '1×1, 32' represents that in this convolution layer, kernel size and the filter numbers are 1×1 and 32, respectively.

The multi-scale sequence feature maps are then fed into Part B after the ReLU function has acted on them. Part B mainly performs up sampling and down sampling operations with the aim of ensuring that the feature maps at different scales have the same number of feature channels before they are fused. The up sampling first passes through a layer of Conv1D and then performs an Upsanmpling1D operation, where the number of convolution filters in Conv1D is the same as the size of the corresponding input multi-scale feature map, with kernel size of 1×1, stride of 1 and an Upsanmpling1D size of 2. The down sampling consists of only one layer of Conv1D, with kernel size of 9×1 and strides of 2. The number of convolutional filters is the same as for the up sampling, and eventually the different scales of the corresponding channels are summed to obtain a sequence of fused feature maps.

The fused sequence feature maps are input to Part C after the functions of the ReLU and Lambda, which firstly input feature maps into a residual layer of Conv1D with kernel size of 9×1 and strides of 2, and the number of convolution filters corresponds to the size of the corresponding input fused feature maps, then followed by a layer of Conv1D with kernel size of 1×1, stride of 1 and the number of convolutional filters is twice the first dimension of the corresponding output feature map tensor, and then following a layer of one-dimensional global averaging pooling to remove redundant information and reduce the size of the feature maps. The final fully-connected layer is used to complete the classification and which contains 24 cells. To normalize the output of each cell, the fully-connected layer uses the Softmax function as the activation function and the output is the probability that the target signal belongs to the corresponding modulation type. This process can be represented as:

$$\hat{y} = arg\ max(soft\ max(x_f)) \quad (8)$$

Where $\hat{y}$ denotes the predictions obtained and $x_f$ denotes the output features of the fully connected layer.

Moreover, as shown in Fig. 2. Almost every Conv1D convolutional filter of SE-MSFN is used in combination with batch normalization and ReLU activation except the last fully connected layer, which can improve the convergence speed and accuracy of the network during the training phase.

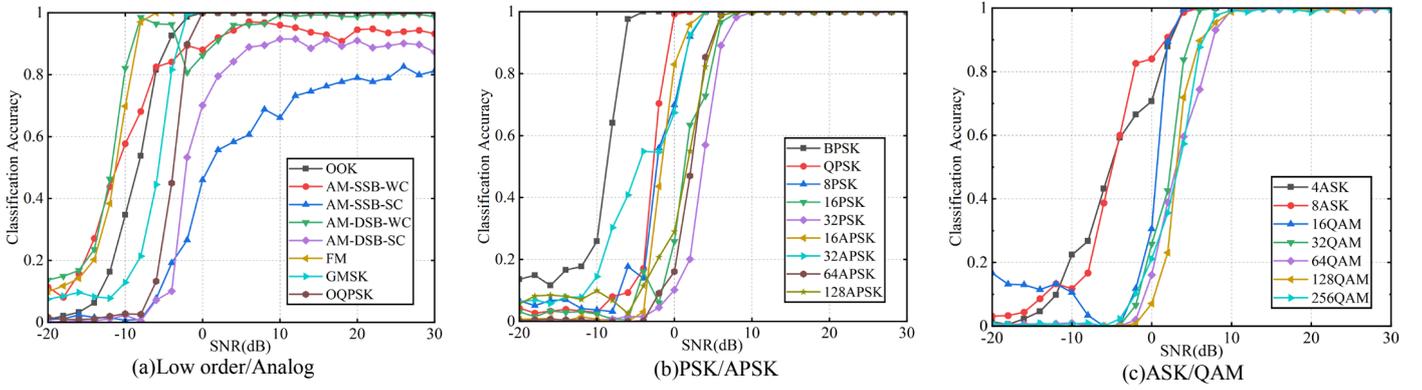

Fig. 6. Classification accuracy of SE-MSFN under different modulation types. With the sequences: (a) Low order/Analog, (b) PSK/APSK, and (c) ASK/QAM.

*D. Ensemble learning*

The idea of ensemble learning is principally based on the theory foundation stone that the generalization ability of an ensemble is usually much stronger than that of a single learner [35]. AMC is a supervised learning classification task and the different hyper-parameter combinations of SE-MSFN will result in different classification performance, if the strengths of each model can be exploited effectively, it will work well. So, the method of ensemble learning can be used to make several sets of relatively weak models into one strong model, the trained several base models are first predicted by all base models in the prediction phase to form a new test set, and finally the voting is taken to predict the test set again.

## V. EXPERIMENT

*A. Experiment environment and Dataset*

The model was built using the Keras framework embedded in TensorFlow2.2 under the Ubantu18.04.5 operating system, and the algorithm was implemented on an NVIDIA TiTan RTX×2 GPU. The network was trained from scratch in 200 epochs, the batch size for each iteration is set to 3000. We trained the network using Adam as the optimizer with a learning rate of 0.001. As shown in Table I. The dataset is RADIOML 2018.01A, which contains 24 common analog and digital modulated modes, we divided the dataset into a training set, a verification set and a test set according to a ratio of 8:1:1.

Table I. THE DETAILS OF DATASET

| Dataset name | RADIOML 2018.01A |
|---|---|
| Modulation modes | OOK, 4ASK, 8ASK, BPSK, QPSK, 8PSK, 16PSK, 32PSK, 16APSK, 32APSK, 64APSK, 128APSK, 16QAM, 32QAM, 64QAM, 128QAM, 256QAM, AM-SSB-WC, AM-SSB-SC, AM-DSB-WC, AM-DSB-SC, FM, GMSK, OQPSK |
| Signal dimension of per sample | 1024×2 |
| SNR(dB) | -20:2:30 |
| Number of samples | 2559604 |

*B. Experiment results*

Fig. 6 shows the classification performance of SE-MSFN, and we have divided the 24 modulation modes into three types for ease of observation. When the SNR is higher than 10dB, almost all types of modulation achieve close to 100% classification accuracy except four analog modulation modes e.g. AM-SSB-WC, AM-SSB-SC, AM-DSB-WC, AM-DSB-SC. In particular, the higher-order APSK and QAM which are not easily classified, achieve nearly 100% classification accuracy for high SNR $\geq$ 10dB. When the SNR is in the range of 0-10dB, five modulation modes e.g. OOK, FM, GMSK, 16QAM and BPSK can achieve 100% classification accuracy, and eight modulation modes e.g. 4ASK, 8ASK, QPSK, 8PSK, 16APSK, 32APSK, AM-DSB-WC and AM-SSB-WC can reach an average classification accuracy of 93%. More than 60% of modulation modes achieve an average classification accuracy of above 83%. When the SNR is below 0dB, the classification accuracy of all modulation modes is generally lower due to the worse conditions of the communication channel.

Fig. 7 shows the confusion matrix for 24 modulation modes at SNR of +10dB. which facilitates detailed analysis of classifier performance, most of modulation modes can be accurately classified and less confused, but AM-SSB-SC confused with AM-SSB-WC, AM-DSB-WC confused with AM-SSB-WC. The reason is that they all contain AM modulation modes and have extremely similar features, making it difficult for network to classify them well. To solve it, maybe we can adopt a two-stage classification strategy. Feeding these four analogue signals into an additional network for re-judgement, which can make up for the shortcomings of the primary network.

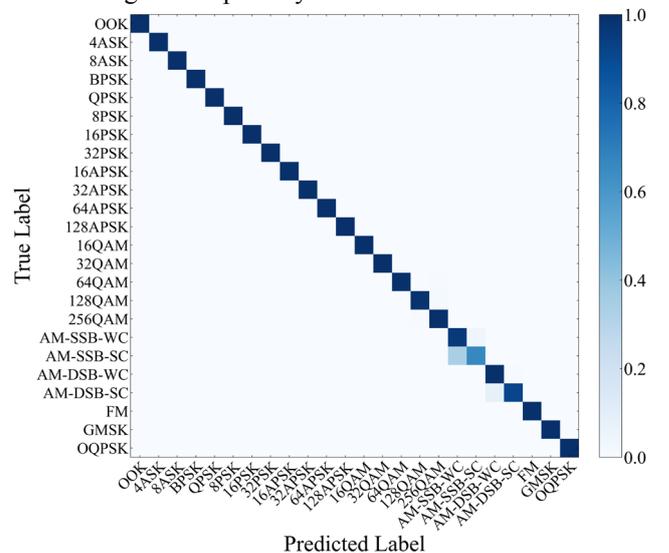

Fig. 7. Confusion matrix of 24-modulation classification at +10dB SNR

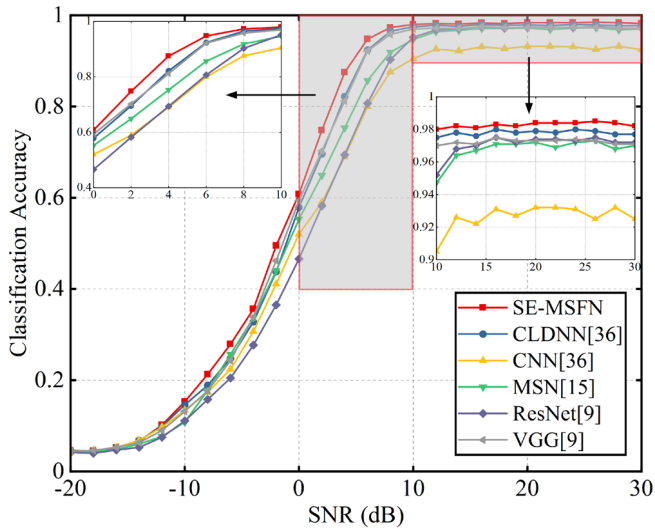

Fig. 8. Comparison of overall accuracy with other methods at block of 4, reduction ratio of 1, repetition of 2 and kernel size of 9.

Table II. COMPARISON OF CLASSIFICATION ACCURACY AND COMPUTATIONAL COMPLEXITY WITH OTHER METHODS

| Network | Average Accuracy (%) | | Inference Time(ms) | Capacity (No. parameters) |
|---|---|---|---|---|
| | *All SNR* | *SNR of 0-10dB* | | |
| **ResNet [13]** | 59.62 | 73.40 | 0.051 | 205K |
| **VGG [13]** | 60.48 | 76.83 | 0.043 | 283K |
| **MSN [19]** | 61.60 | 77.97 | 0.128 | 143K |
| **CLDNN[36]** | 63.08 | 82.68 | 0.052 | 847K |
| **CNN [36]** | 58.53 | 73.02 | 0.044 | 466K |
| **SE-MSFN** | **64.50** | **85.53** | 0.137 | 332K |

Fig. 8 compares the classification performance of six models, including CLDNN [36], CNN [36], MSN [19], ResNet [13] and VGG [13]. As shown in Table II. The average classification accuracy of SE-MSFN is higher than the other methods, surpassing CLDNN [36] by 1.42% and MSN [19] by 2.9%. In particular, for the range of SNR is 0 to 10dB, which is closest to the real communication channel environment, the average classification accuracy of SE-MSFN far exceeds that of other models, surpassing CLDNN [36] by 2.85% and MSN [19] by 7.56%. When SNR is in the range of -4 to 4 dB, the average classification accuracy of SE-MSFN is 61.64%, which surpasses CLDNN [36] and MSN [19] by 4.38% and 6.97%, respectively. Which indicates that our model can better extract the in-depth features of the signal has at lower SNR. Comparing the computational complexity, the parameter number of SE-MSFN is much smaller than that of CLDNN [36] and CNN [36]. Although our model has a medium level of computational complexity among all models and takes a slightly longer time to predict, we achieve better classification performance.

As can be seen from the Fig. 9.The training process of SE-MSFN is smoother than CLDNN [36] and MSN [19], and with almost no oscillation. Also, it takes less time to achieve the same classification accuracy as CLDNN [36] and MSN [19]. Which indicates that the SE-MSFN converges faster than other methods.

SE-MSFN has four key hyper-parameters, i.e. kernel size, block, reduction ratio, and repetition. The kernel size represents the size of the Conv1D used to extract the feature

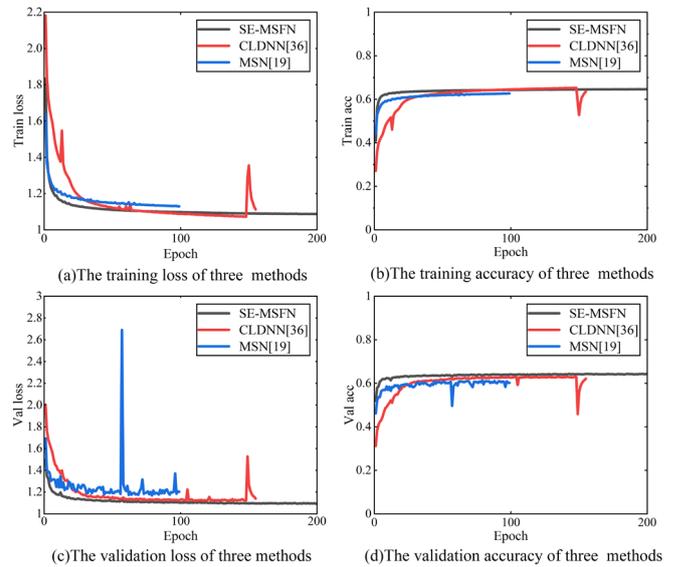

Fig. 9. The training loss(a), train accuracy(b), validation loss(c), and validation accuracy(d) data of three deep learning methods on RADIOML 2018.01A of the raw signal I/Q sequence data. The block represents the number of bottleneck blocks. The reduction ratio determines the number of convolutional filters in the first fully connected layer of the SE Module, and the repetition represents the number of Residual layer. First of all, we tested the classification performance and computational complexity of the model with different combinations of the other three key hyper-parameters at kernel size fixed to 8. As shown in Table III. The best classification performance is achieved when the combination of hyper-parameters is block of 4, reduction ratio of 1 and repetition of 2, which with an average classification accuracy of 64.41% and maximum classification accuracy of 98.3%.

By comparing the experimental results of the corresponding rows of Table III using the method of controlling variables. We can analyze that when the other two hyper-parameters are kept constant, the classification accuracy increases with the increasing of block and repetition for the reason that the more bottleneck blocks are stacked on top of each other, the better extraction of multi-scale feature maps will be. Also with the help of the SE module, the feature extraction module can further extract the more important multi-scale features in the signal, and at the cost of higher computational complexity, as more convolutional operations are required. The last four rows (rows 13 to 16) of the Table III illustrate that the reduction ratio has little effect on the classification accuracy and computational complexity of the model. The reason is that when SE Module performs the operation of excitation, the output of the first fully-connected layer is directly input to the second fully-connected layer after the function of ReLU, and then the weights of each channel are obtained by the sigmoid function, the first fully-connected layer mainly plays the role of raising the dimension, which has less influence compared to the second fully-connected layer. Furthermore, the maximum classification accuracy of all hyper-parametric combinations are above 98% except for the for the two simplest parameter combinations in rows 11 and 12. which indicates that the proposed method is more robust to parameter variations at high SNR.

Table III. COMPARISON OF MODEL RECOGNITION PERFORMANCE WITH DIFFERENT COMBINATIONS OF HYPERPARAMETERS.

| NO. | Block | Reduction Ratio | Repetition | Average Accuracy (%) | Maximum Accuracy (%) | Train Time(s/epoch) | Inference Time(ms) | Capacity (No. parameters) |
|---|---|---|---|---|---|---|---|---|
| 1 | 4 | 1 | 3 | 64.22 | 98.3 | 614 | 0.206 | 1452K |
| 2 | 3 | 1 | 3 | 64.16 | 98.2 | 523 | 0.171 | 1191K |
| 3 | 2 | 1 | 3 | 64.12 | 98.3 | 411 | 0.144 | 929K |
| 4 | 1 | 1 | 3 | 63.72 | 98.2 | 312 | 0.117 | 667K |
| 5 | **4** | **1** | **2** | **64.41** | **98.3** | 364 | 0.132 | 307K |
| 6 | 3 | 1 | 2 | 63.97 | 98.4 | 304 | 0.113 | 244K |
| 7 | 2 | 1 | 2 | 64.05 | 98.2 | 238 | 0.096 | 181K |
| 8 | 1 | 1 | 2 | 63.31 | 98.0 | 169 | 0.088 | 118K |
| 9 | 4 | 1 | 1 | 63.30 | 98.1 | 194 | 0.082 | 64K |
| 10 | 3 | 1 | 1 | 63.18 | 98.0 | 163 | 0.074 | 51K |
| 11 | 2 | 1 | 1 | 62.50 | 97.9 | 131 | 0.064 | 39K |
| 12 | 1 | 1 | 1 | 60.53 | 96.0 | 99 | 0.054 | 26K |
| 13 | 4 | 4 | 2 | 64.27 | 98.4 | 614 | 0.136 | 276K |
| 14 | 4 | 8 | 2 | 64.10 | 98.3 | 358 | 0.145 | 271K |
| 15 | 4 | 12 | 2 | 64.15 | 98.5 | 357 | 0.145 | 269K |
| 16 | 4 | 16 | 2 | 64.18 | 98.3 | 358 | 0.131 | 268K |

Table IV. CLASSIFICATION ACCURACY AND COMPUTATIONAL COMPLEXITY OF SE-MSFN UNDER DIFFERENT KERNLE SIZE.

| Kernel Size | Average Accuracy (%) | Maximum Accuracy (%) | Inference Time(ms) | Capacity (No. parameters) |
|---|---|---|---|---|
| 3 | 63.68 | 98.2 | 0.123 | 178k |
| 5 | 63.82 | 98.1 | 0.128 | 229k |
| 7 | 64.30 | 98.4 | 0.131 | 281k |
| 8 | 64.41 | 98.3 | 0.132 | 306k |
| 9 | **64.50** | **98.5** | 0.136 | 332k |
| 11 | 64.38 | 98.4 | 0.145 | 383k |
| 13 | 64.43 | 98.4 | 0.150 | 435k |
| 15 | 64.50 | 98.4 | 0.161 | 486k |

The average classification accuracy of SE-MSFN with different kernel size are shown in Table IV. We first tried the odd convolutional kernel sizes from 3 to 15 in order to avoid possible alignment errors in the process of convolutional operations with even convolutional kernel sizes [37]. However, we found that the classification performance of the model was not affected by that problem for the reason that the global features of the time series data are more important than the boundary features. When kernel size ≤ 9, classification accuracy increases with increasing convolutional kernel size. When kernel size ≥ 9, the classification accuracy doesn't continue to rise but still remained above 64%. The reason is that the effective sensory field of the large kernel size is more beneficial to extract global features of the sequence data, can also focus on the backward and forward correlation and long-term dependence of the time series data more effectively. Ultimately, we set the convolution kernel size of SE-MSFN to 9.

Table V. THE COMPARISON OF CLASSIFICATION ACCURACY AND COMPUTATIONAL COMPLEXITY BETWEEN SE-MSFN AND MSFN WITHOUT SE.

| Network | Average Accuracy (%) | Maximum Accuracy (%) | Inference Time(ms) | Capacity (No. parameters) |
|---|---|---|---|---|
| Without SE | 62.82 | 97.5 | 0.134 | 291K |
| **SE-MSFN** | **64.50** | **98.5** | 0.137 | 332K |

As shown in Table V. The classification performance of the model is significantly reduced without SE module for the reason that the bottleneck blocks are unable to extract important features of the signal without channel self-attention mechanism.

Table VI. COMPARISON OF CLASSIFICATION PERFORMANCE AFTER ENSEMBLE LEARNING. (The text '412_ks=9' means 'block=4, reduction ratio=1, repetition=2, and kernel size=9', text 'EN' means 'Ensemble'.)

| Network | | Average Accuracy(%) | | Maximum Accuracy(%) | Inference Time(ms) |
|---|---|---|---|---|---|
| | | *All SNR* | *SNR of 0-10dB* | | |
| **412_ks=9** | | 64.50 | 85.53 | 98.5 | 0.137 |
| En | 412_ks=9 | **64.86** | **86.3** | **98.6** | 0.796 |
| | 412_ks=13 | | | | |
| | 412_ks=15 | | | | |

Finally, as shown in Table VI. The use of ensemble learning can help further improve the classification accuracy of the three models (kernel size = 9, 13, and 15), but at the cost of a longer inference time.

VI. CONCLUSION

This paper proposed a multi-scale convolutional neural network AMC method named SE-MSFN, which adopts a large size convolutional filter design to better focus on the correlation and dependency of the raw signal I/Q sequence data, and incorporates the SE self-attentive mechanism into the network to extract more important features of the signal. The proposed method achieves an average classification accuracy of 64.5%, a maximum classification accuracy of 98.5% and an average classification accuracy of 85.53% in the lower SNR range 0 to 10dB, which is better than other existing methods. In addition, ensemble learning can help further improve the classification accuracy. Finally, we hope that the research in this paper can provide some reference for engineers and researchers in this field in terms of engineering practice and experimental research.